\documentclass[twocolumn]{aastex6}
\usepackage{epsfig}
\usepackage{natbib}
\usepackage{amsmath}
\usepackage{color}
\usepackage{graphicx}
\usepackage[caption=false]{subfig}

\newcommand{\kms}{km~s$^{-1}$}

\newcommand{\kmsMpc}{km~s$^{-1}$~Mpc$^{-1}$}

\DeclareMathAlphabet{\mathpzc}{OT1}{pzc}{m}{it}

\begin{document}
\title{Cosmicflows-3: Cosmography of the Local Void}

\author{R. Brent Tully,}
\affil{Institute for Astronomy, University of Hawaii, 2680 Woodlawn Drive, Honolulu, HI 96822, USA}
%\and
\author{Daniel Pomar\`ede}
\affil{Institut de Recherche sur les Lois Fondamentales de l'Univers, CEA, Universite' Paris-Saclay, 91191 Gif-sur-Yvette, France}
%\and
\author{Romain Graziani}
\affil{University of Lyon, UCB Lyon 1, CNRS/IN2P3, IPN Lyon, France}
%\and
\author{H\'el\`ene M. Courtois}
\affil{University of Lyon, UCB Lyon 1, CNRS/IN2P3, IPN Lyon, France}
%\and
\author{Yehuda Hoffman}
\affil{Racah Institute of Physics, Hebrew University, Jerusalem, 91904 Israel}
%\and
\author{Edward J. Shaya}
\affil{University of Maryland, Astronomy Department, College Park, MD 20743, USA}

\begin{abstract}
{\it Cosmicflows-3} distances and inferred peculiar velocities of galaxies have permitted the reconstruction of the structure of over and under densities within the volume extending to $0.05c$.  This study focuses on the under dense regions, particularly the Local Void that lies largely in the zone of obscuration and consequently has received limited attention.  Major over dense structures that bound the Local Void are the Perseus-Pisces and Norma-Pavo-Indus filaments separated by 8,500~\kms.  The void network of the universe is interconnected and void passages are found from the Local Void to the adjacent very large Hercules and Sculptor voids.  Minor filaments course through voids.  A particularly interesting example connects the Virgo and Perseus clusters, with several substantial galaxies found along the chain in the depths of the Local Void.  The Local Void has a substantial dynamical effect, causing a deviant motion of the Local Group of $200-250$~\kms.  The combined perturbations due to repulsion from the Local Void and attraction toward the Virgo Cluster account for $\sim50\%$ of the motion of the Local Group in the rest frame given by the cosmic microwave background.

\bigskip
\end{abstract}

\smallskip
\section{Introduction}
The average place in the universe is in a void.  The Local Void \citep{1987nga..book.....T} subtends 40\% of the sky and begins 1 Mpc away, at the fringe of the Local Group.  Over the eons, matter evacuates from voids and builds up in adjacent sheets, filaments, and knots, the components of the cosmic web \citep{1996Natur.380..603B}.   Most of the matter that makes up our galaxy and that of our neighbors must have come out of the Local Void so our relationship to that structure is fundamental to attempts to understand details of the local neighborhood \citep{2013MNRAS.436.2096S, 2016MNRAS.458..900C}.  

There is increasingly good information about the kinematics of nearby galaxies from distance measurements using the tip of the red giant branch technique that conclusively demonstrates the motions of galaxies away from the Local Void  \citep{2015ApJ...805..144K, 2017ApJ...835...78R, 2017ApJ...850..207S, 2018AJ....156..105A}.  Studies of the nearby region provide a unique opportunity: only nearby are deviant velocities comparable to cosmic expansion velocities to the degree that these motions can be cleanly separated in individual cases.  So motions are observed consistent with expansion of the Local Void.  Are these motions of an amplitude that theory would anticipate?

The Local Void has been difficult to study because it is located behind the center of the Milky Way.  It is so large that it easily protrudes on both sides of the galactic plane, but much of it is obscured.  This paper gives attention to a way to study the morphology of the Local Void that is relatively insensitive to direct observation.  {\it Cosmicflows-3} (CF3) is a collection of 18,000 galaxy distances \citep{2016AJ....152...50T} that, although deficient in the zone of obscuration, captures the essence of structure all-sky through two alternative analyses.
Both analyses assumes that structure forms from Gaussian initial fluctuations within a $\Lambda$ Cold Dark Matter universe with matter and energy densities characterized by $\Omega_m=0.3$, $\Omega_{\Lambda}=0.7$.  

One method involves Wiener filtering with constrained realizations \citep{1999ApJ...520..413Z, 2012ApJ...744...43C} and is the methodology used in previous Cosmicflows papers \citep{2014Natur.513...71T}.
Within the $\Lambda$CDM paradigm and the linear approximation, the Wiener Filter provides the optimal Bayesian estimator of the confluence of the linear growth of the assumed power spectrum of perturbations and the observed constraints \citep{1995ApJ...449..446Z}.  The observed constraints are peculiar velocity estimates, $V_{pec}$, derived from distance measurements, $d$, where to first approximation peculiar velocities are decoupled from observed velocities, $V_{obs}$, as $V_{pec} = V_{obs} - H_0 d$, with $H_0$ the value of the Hubble Constant consistent with the ensemble of the data.   With the current collection of distances the appropriate value is $H_0 = 75$~\kmsMpc. The direct products are the three-dimensional velocity field and associated density field in the linear regime.

The other method, found to be compatible with the Wiener Filter procedure and used in the model described in this paper, follows the work by \citet{2016MNRAS.457..172L} and is described in detail by \citet{2019arXiv190101818G}.  In simple terms, peculiar velocities imply a distribution of density perturbations that, in turn, imply a velocity field.  A Bayesian procedure is used to estimate the posterior probability of a specific velocity field given the linear theory relationship between densities and velocities through the observed distances with assigned errors.

In addition to constraints on the velocity field and correlated distances, the model solves for a velocity dispersion parameter, $\sigma_{NL}$, that accommodates departures from linear theory, and an effective Hubble Constant.  The model begins with a fiducial value of $H_0 = 75$~\kmsMpc\ \citep{2016AJ....152...50T} but searches for the optimum of a parameter $h_{eff}$ anticipated to be near unity (whence $H_0 = 75 h_{eff}$).  There are uncertainties in both velocities and distances.  Those on velocities are relatively minor and are approximated by $\sigma_{cz} = 50$~\kms.  The errors in distances, in the modulus, are much more substantial.  In recognition that {\it Cosmicflows-3} is a heterogeneous collection of distances, \citet{2019arXiv190101818G} give separation to five sub-samples with each one described by a distinct selection function.

A model that abides by these constraints is sampled by the Markov Chain Monte Carlo (MCMC) method of the Gibbs sampling algorithm \citep{2016MNRAS.457..172L}, whereby each free parameter is drawn from its conditional probability given specification of the other parameters.  The procedure is described in detail by \citet{2019arXiv190101818G} but in brief: 
(a) the parameter $h_{eff}$ is sampled, marginalized over the velocity field;
(b) the conditional probability of the parameter $\sigma_{NL}$ is drawn with the other parameters fixed;
(c) a constrained realization of the density field is drawn assuming a $\Lambda$CDM power spectrum \citep{1991ApJ...380L...5H};
(d) a new set of distances is established from the sampled constrained realization with probabilities set by the current values of $h_{eff}$, $\sigma_{NL}$, and the velocity field, within priors on the distances.  The process is carried through $\sim 10^3$ MCMC steps until convergence.  The procedure has been carried out on multiple constrained realizations and mock catalogs.  With the current analysis, \citet{2019arXiv190101818G} find $h_{eff} = 1.02\pm0.01$ and $\sigma_{NL} = 280\pm35$~\kms.

The resultant model makes predictions for the morphology and motions of structure locally within the $\Lambda$CDM framework and linear perturbations.  
Our present interest is in voids.
It will be asked to what degree the overall model is in agreement with the excellent knowledge we have of the motions of very nearby galaxies.

\section{Morphologies of Nearest Voids}

The Local Void does not have a simple shape.  Moreover, as the void is followed to shallower levels it merges with adjacent voids, as part of a continuous network that extends beyond the volume that can currently be mapped.  The three-dimensional interplay between complex high and low density structures is visually confusing.  We should not have the ambition to get into great detail.

As a prelude, previous efforts to identify nearby voids can be mentioned, derived from regions of emptiness in maps of the distribution of galaxies in redshift surveys. Among the earliest were the seminal studies of the Bo\"otes Void by \citet{1981ApJ...248L..57K} and the void in front of the "Great Wall" Coma and Abell~1367 clusters \citep{1978ApJ...222..784G}.  On very large scales there is the pioneering work by \citet{1985AJ.....90.1413B} and \citet{1994MNRAS.269..301E} on the concentrations and absences of rich clusters.  More nearby and pertaining to the distribution of individual galaxies, of note is the work of \citet{1991MNRAS.248..313K} and  \citet{1998lssu.conf.....F} who, in the latter reference gives a list of 33 void-like regions within 8,500~\kms.  \citet{2013AstBu..68....1E} have produced a more quantitatively rigorous catalog of 89 voids within 3,000~\kms; spherical regions with no known galaxies brighter than $M_K=-18.4$.  Typically these entities are modest in size with radii $\sim 6$~Mpc.   It is well documented that voids network, and their dimensions as constrained by the exclusion of galaxies depends on the intrinsic properties of the galaxy samples.  Sparse filaments of dwarf galaxies can snake through regions devoid of bright galaxies.  This phenomenon has particularly been noted by \citet{1995A&A...301..329L} and \citet{2019MNRAS.482.4329P} within the volume that attracts our attention.  The linkage of the region under consideration including the Local Void to very extensive voids has been claimed by  \citet{2016MNRAS.462.1882K}.  They claim a connection to putative huge voids in the direction of the Cold Spot seen in the temperature fluctuation map of the cosmic microwave background \citep{2015MNRAS.450..288S, 2016MNRAS.455.1246F}.  We find support for this general claim from a large scale flow pattern in our velocity reconstruction based on CF3 distances \citep{2017ApJ...847L...6C}.

The definition of voids based on the distribution of observed galaxies faces serious challenges.  For one, redshift surveys are flux limited which means the back sides of voids are more poorly delineated than the front sides.  In the case of very big voids, the sorts that interest us, this degradation of knowledge with distance is severe.  Survey edge effects is a related concern.  Big voids spill into, and get lost, in the zone of galactic obscuration.  Also, for simply technical reasons, redshift surveys may not provide uniform all-sky coverage, inconveniently clipping potential areas of interest.  Then, it is perhaps the worst of problems that galaxy surveys provide only sparse coverage.  Unlike in simulations where structure can be represented by large numbers of particles, the structure as mapped by individual galaxies is inevitably paltry.

Alternatively, the inhomogeneous distribution of matter can be recovered from the measurement of galaxy distances and the inferences of peculiar velocities.  Galaxies are test particles sampling the gravitational potential.  \citet{2017NatAs...1E..36H} demonstrated the importance of large voids on flow patterns. 

The discussion will make references to a video accompaniment.\footnote{https://vimeo.com/326346346/35088b5dd8} and to two interactive models.\footnote{https://sketchfab.com/models/f0a44df256aa4faf93391887d66010e2 and https://sketchfab.com/models/78885b3d303d4b6e99cfe099b43929fb}
The complex three-dimensional nature of large scale structure is most easily dissected with the capabilities of zoom and motion of a movie and interactive models.

\subsection{Local Void}

\begin{figure*}[]
%\begin{center}
%\special{psfile=overview.ps hscale=50 vscale=50 voffset=-100 hoffset=0}
%\includegraphics[width=0.4\textwidth]{overview.ps}
%\plotone{overview.ps}
\plotone{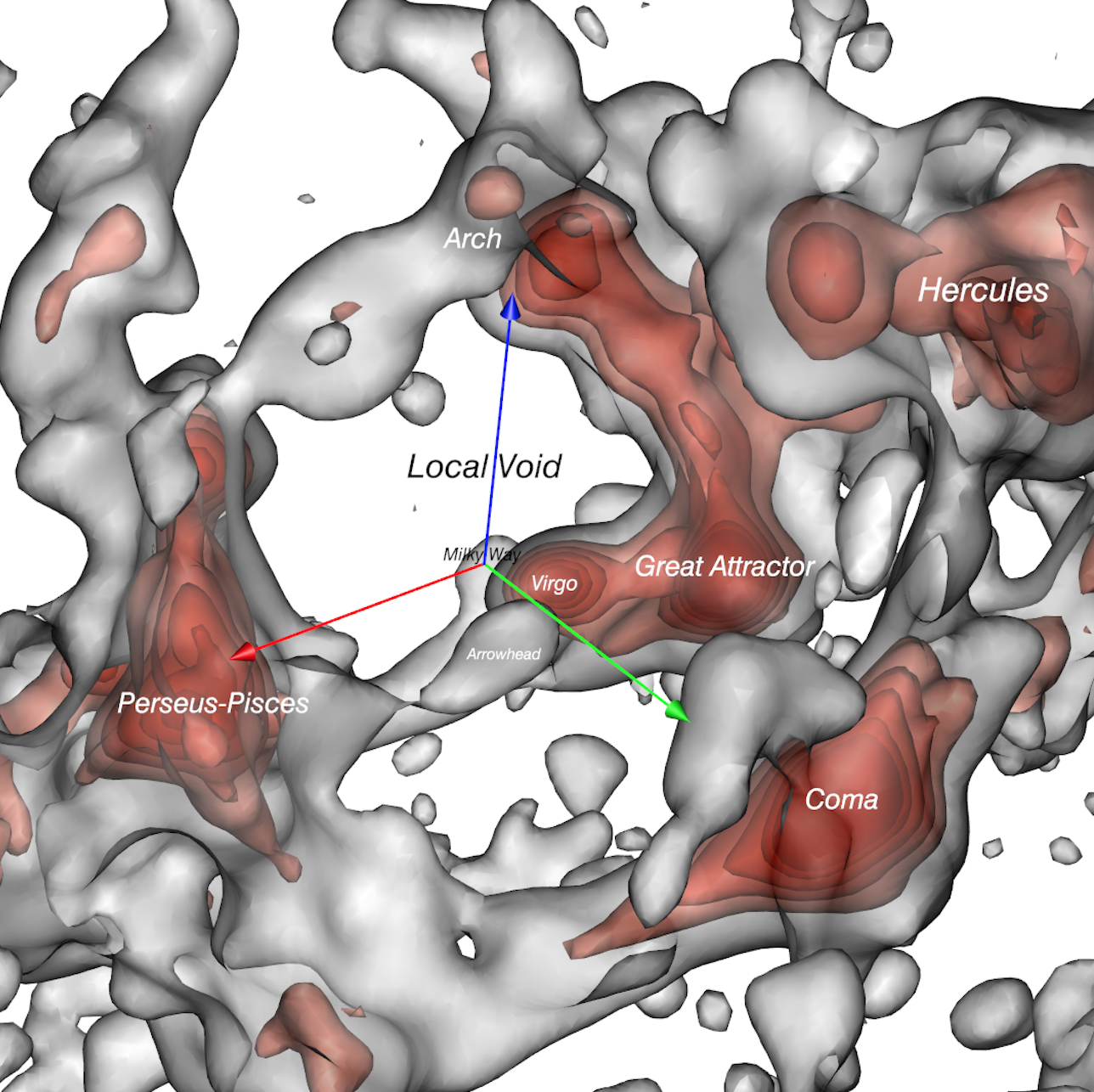}
\caption{Overview of the structure surrounding the Local Void.  Isosurfaces of density are inferred from the velocity field constructed from the Wiener Filter treatment of Cosmicflows distances, with the densest peaks in red and less dense filaments in grey.  The Milky Way is at the origin of the colored arrows, 5,000~\kms\ in length, oriented in the frame of supergalactic coordinates (red toward +SGX, green toward +SGY, blue toward +SGZ).  The Local Void fills the empty region above the Milky Way in this plot.
This view inward from a location at positive values of SGX, SGY, and SGZ will be referred to as the reference orientation.
}
%\end{center}
\label{overview}
\end{figure*}

Reigning in the focus to nearby, consider the structure represented in Figure~\ref{overview}.  Here we see a smoothed description of over dense regions in our vicinity extending to $\sim 10,000$~\kms.  The Local Sheet with our Milky Way at the origin of the plot lies at a density less than the lowest grey contour.  Major knots are identified: the Virgo Cluster, the Perseus-Pisces filament \citep{1988lsmu.book...31H}, the Coma Cluster within the Great Wall climbing to the Hercules complex \citep{1986ApJ...302L...1D} and, nearer home, the Great Attractor region \citep{1987ApJ...313L..37D} with the Pavo-Indus filament rising above it connecting to a feature we call the Arch \citep{2017ApJ...845...55P} that caps the Local Void and provides a connection to Perseus-Pisces. 

The structure shown in Figure~\ref{overview} is entirely derived from an analysis of departures from cosmic expansion from samples of galaxies with measured distances.  The specific rendition shown in this figure is extracted from the quasi-linear construction described by \citet{2018NatAs...2..680H}.  Thanks to the large scale coherence of velocity flows, loss of information in the zone of obscuration has minimal impact on the derived model and features are robust within $\sim 8,000$~\kms\ where the density of test particles with distance measures is high.

With the upper left panel of Figure~\ref{LV}, the same reference perspective is preserved but we move in closer.  Here and in following figures unless explicitly stated, the layered surfaces are density iso-contours of the \citet{2019arXiv190101818G} reconstruction derived from {\it Cosmicflows-3} distances. Over or under densities, $\delta({\bf r})$, follow from the gradient of velocities, ${\bf v}$, in linear theory:
\begin{equation}
\delta({\bf r}) = -\nabla\cdot{\bf v} / H_0 f
\end{equation} 
where $f$ is the growth rate of structure assuming standard $\Lambda$CDM parameters.  The over density surfaces begin at $\delta = 0.75$ in grey and progress through increasingly strong shades of red with $\delta$ levels 1.00, 1.25, 1.50, 1.75, 2.00, 2.25.  The under dense levels are $-0.7$ and $-1.1$ with two levels, and a shallower $-0.2$ if a third level is shown.

In Figure~\ref{LV}, the core of the Local Void is represented at two density contours of black and dark grey.  The panels show the same scene from different vantage points, as specified in the figure caption.  The overdense contours are stripped away in the lower right panel to fully reveal the Local Void.

  We introduce a naming convention that will be adhered to in subsequent figures.
The names of familiar structures are retained.
Otherwise, features are given constellation names appended with a tag set by their redshift in units of $10^3$~\kms, with the tags of under densities preceded by a minus sign and those of over densities preceded by a plus sign.   
Here in the Local Void, Lacerta$-2.4$ is at the location of the lowest density of $-1.89$ at supergalactic SGX, SGY, SGZ of [+1650, $-700$, +1650]~\kms\ $\approx$ [+22, $-9$, +22]~Mpc.  
Andromeda$-2.3$ is at a secondary minimum of $-1.53$ at [+2100, $-700$, $-300$]~\kms\ $\approx$ [+28, $-9$, $-4$]~Mpc and, in the most familiar part of the Local Void, Aquila$-0.8$ is a tertiary minimum  of $-1.13$ at SGX, SGY, SGZ of [$-200$, $-200$, 700]~\kms\ $\approx$ [$-3$, $-3$, +9]~Mpc in our immediate vicinity only 10 Mpc away.  
More removed, UMi$-3.7$ marks a minimum of $-0.93$ at [+3100, +1700, +1200]~\kms\ $\approx$[+41, +23, +16]~Mpc.
Details regarding these minima are accumulated in Table~\ref{table:minima}.

\begin{figure*}[]
%\plotone{LV_contours.ps}
%\plotone{LV-fig2test.ps}
%\plotone{LV_4panels.eps}
\plotone{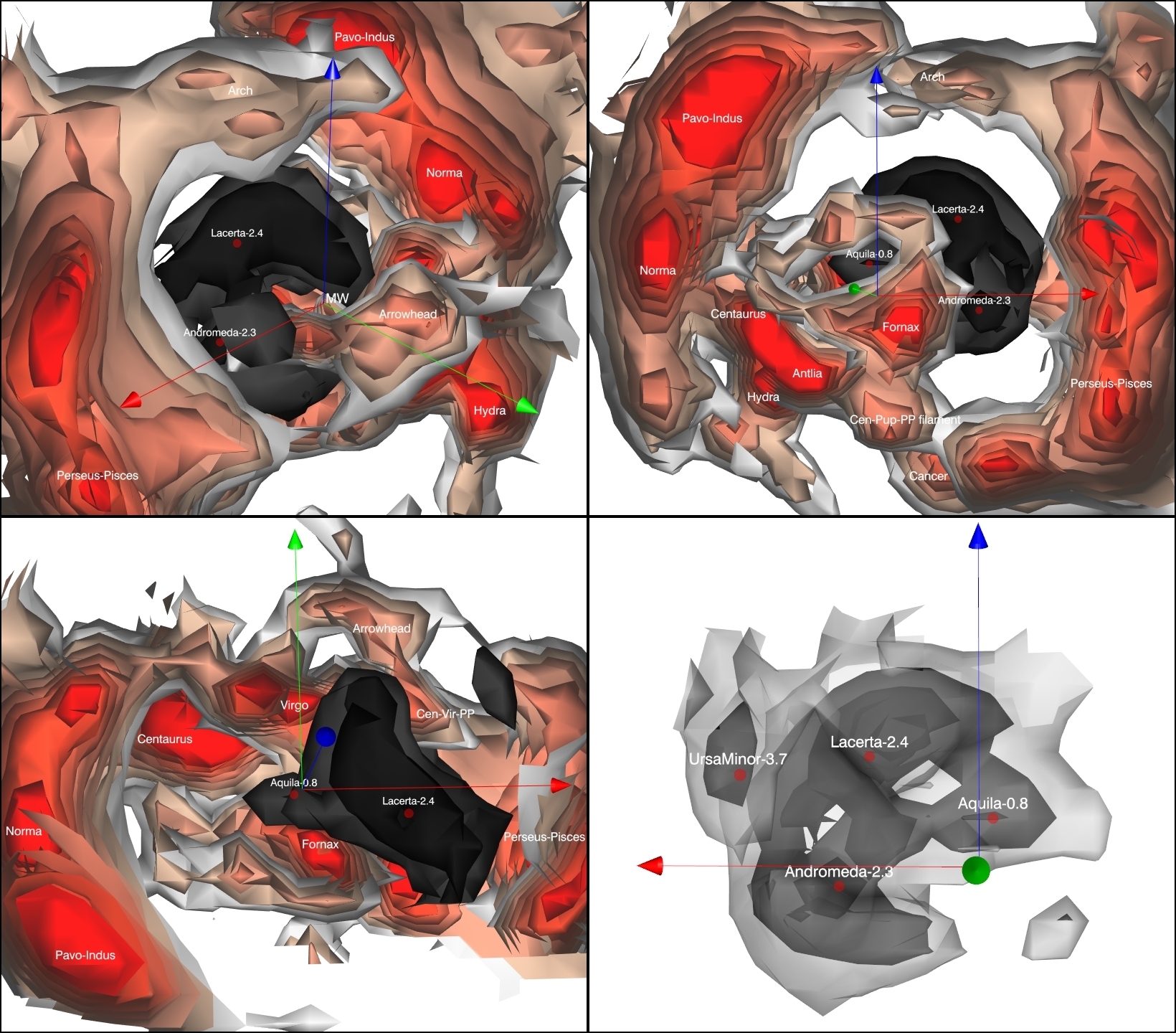}
\caption{The heart of the Local Void.  The deepest parts of the void are mapped by surfaces of density $-1.1$ (black) and $-0.7$ (dark grey).  Local minima are located by red dots and given names.
Contours in shades of light grey and red illustrate surrounding high density structures. The Milky Way is at the origin of the red, green, blue directional arrows.
The same scene is shown from multiple vantage points.  The reference viewing direction in the upper left panel is from positive values of all 3 coordinates (video frame time: 02:01).  At upper right, the scene has been rotated around to almost in from the negative SGY axis (02:25).  Then at lower left, the view is in from very near to the positive SGZ axis.  In this latter case, a foreground clip at SGZ=$+3000$~\kms\ has removed the Arch to give an unrestricted view of the void (02:32). In the lower right panel, the Local Void contours are shown alone, looking in from positive SGY (02:42).
}
\label{LV}
\end{figure*}

The deepest minima in the Local Void lie at very low values of SGY; i.e., they lie close to the equatorial plane of the Milky Way in regions of obscuration.  The void manifests a tilt toward positive SGX, toward the space in front of the Perseus-Pisces filament which is the well documented domain of a void \citep{1986ApJ...306L..55H}.  The CF3 velocity information resolves ambiguity in mapping based on redshift surveys, aggravated by galactic obscuration, and clearly identifies the Local Void and the void foreground of the Perseus-Pisces complex as parts of the same feature.  The "hypervoid" HV1 defined by the union of 56 small spherical voids by \citet{2013AstBu..68....1E} reasonably approximates our Local Void.  The rough dimensions of the Local Void at the isodensity contour $-0.7$ is $\Delta$SGX,SGY,SGZ = 5200,3000,4500~\kms\ = 69,51,60~Mpc, hence a volume of $\sim 2 \times 10^5$~Mpc$^3$.

\begin{figure*}[]
%\plotone{LV_LSCorbits.png}
\plotone{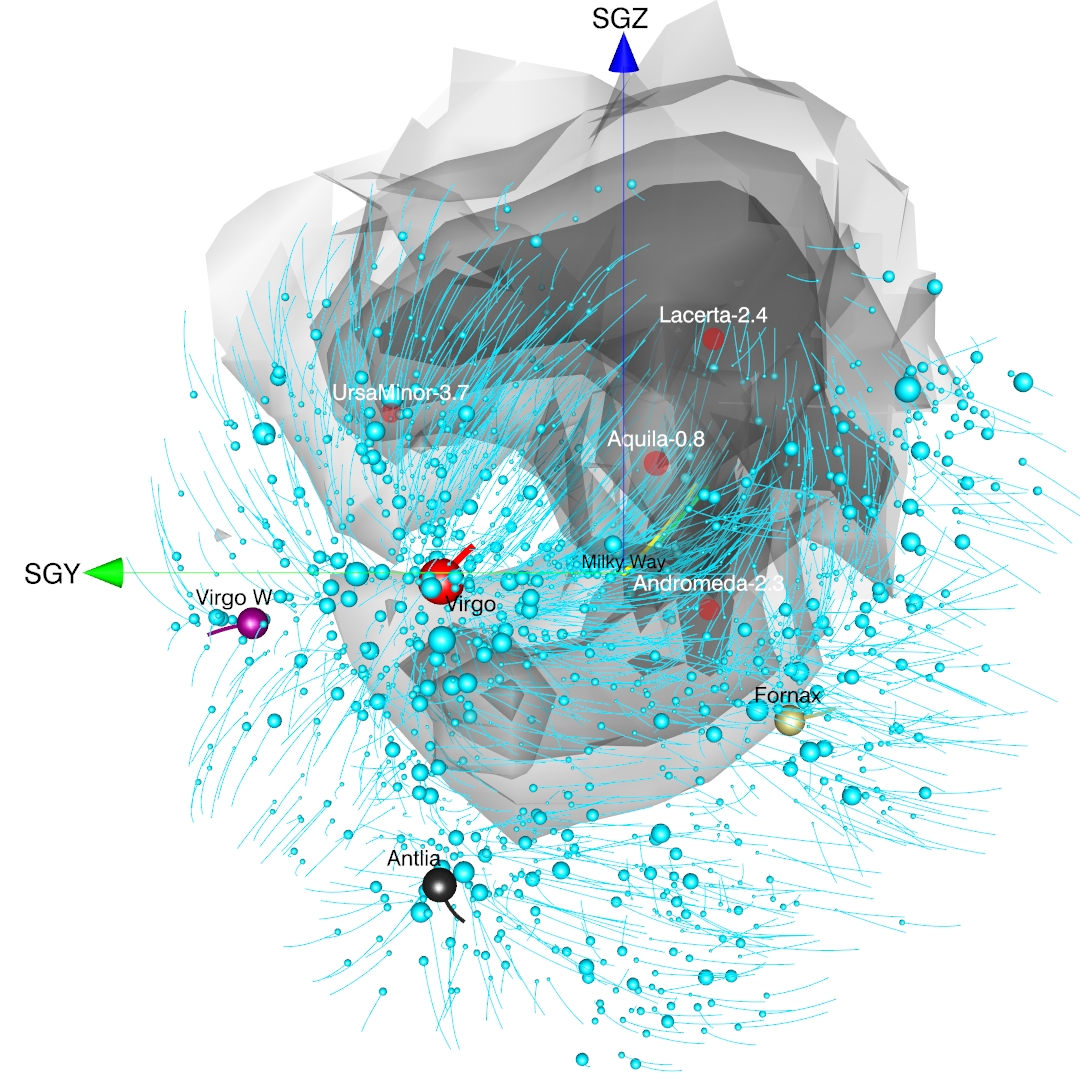}
\caption{
Orbits derived from the numerical action methods of Shaya et al. superimposed on the Local Void iso-density contours.  Orbits systematically descend out of the void (06.11).  In this figure only, the green-blue (SGY-SGZ) coordinate arrows have length 3500 \kms.
}
\label{orbits}
\end{figure*}

A personalized tour of the Local Void stripped of over dense boundaries (Figure~\ref{LV}, lower right panel) can be experienced by accessing the first interactive model.\footnote{https://sketchfab.com/models/f0a44df256aa4faf93391887d66010e2}
The superimposed orbits were derived from {\it Cosmicflows-3} distance constraints using numerical action methods \citep{2017ApJ...850..207S}.    The orbits are calculated in co-moving space coordinates following the center of mass of the sample.   The orbits from $z = 4$ to today dramatically illustrate the evacuation of the Local Void.  See also the sequence in the video frozen in the frame image of Figure~\ref{orbits}.

\subsection{Hercules Void}

\begin{figure*}[]
%\plotone{hercules_void.ps}
%\plotone{HV_2panels.eps}
\plotone{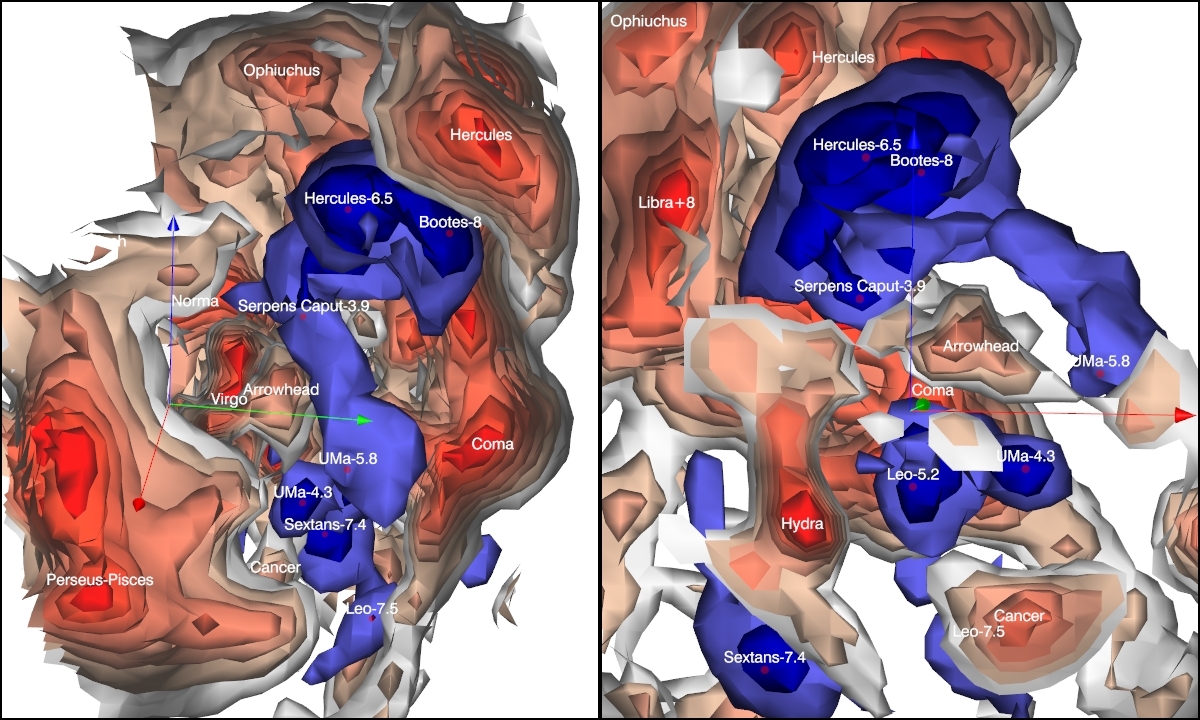}
\caption{The Hercules Void.  The deepest density minimum are shown with contours of blue at the same density levels as in Fig.~\ref{LV}.
The locations of local minima are identified by red dots and names. Major bounding overdensities are identified.
This extended void lies to the foreground of the high density complex of clusters in Hercules.  Multiple lesser extrema lie throughout the extended void that occupies the space from behind the traditional Local Supercluster and Great Attractor complex to the foreground of the Great Wall.
The left panel shows a view from positive SGX and SGZ, slightly rotated from the reference viewing direction (video frame time 07:15).  The view in the right panel is almost along the negative SGY axis, close to the viewing direction in the lower left panel of Fig.~\ref{LV}, but with a foreground clip at SGY=+2200~\kms\ to afford minimal obstruction (07:25). 
}
\label{herc}
\end{figure*}

The Local Void is not isolated from other voids, but before investigating its growth at lesser under densities let us become familiar with the other two principal density depressions within 8,000~\kms.  The one in the north galactic hemisphere (supergalactic SGY$>0$) is seen in Figure~\ref{herc}.  The main part of this entity has been called the Northern Local Void \citep{1983HiA.....6..757E, 1995A&A...301..329L}, an unfortunately confusing name.  \citet{2013AJ....146...69C} refer to the feature as the Hercules Void because of the location of the dominant component directly in front of the Hercules cluster complex.  We use this name.  The deepest minimum in density of $-1.87$ occurs at [$-1200$, +4000, +5000]~\kms\ $\approx$ [$-16$, +53, +67]~Mpc.  A list of secondary minima is given in Table~\ref{table:minima}.  Those called "arrowhead" bracket the Arrowhead mini-supercluster \citep{2015ApJ...812...17P}.  The minima at negative SGZ are parts of what has been called the Southern Local Void \citep{1983HiA.....6..757E}.  These depressions link up as one considers less negative density levels.  Increasing the isodensity cut to more positive values, the entire region behind the traditional Local Supercluster \citep{1953AJ.....58...30D} and in front of the Great Wall \citep{1986ApJ...302L...1D} is revealed to be under dense.  As \citet{1995A&A...301..329L} point out, though, this volume is not devoid of galaxies.  The region is laced with tenuous filaments making connections that span the width of the region.  An example of a filament network connection between the Virgo and Coma clusters is illustrated by \citet{2008AJ....135.1488T}.  The full dimension of what we are calling the Hercules Void at isodensity $-0.7$ is $\Delta$SGX,SGY,SGZ = 11600,7800,14000~\kms\ = 155,104,187~Mpc, a volume of $\sim 3 \times 10^6$~Mpc$^3$.  At the isodensity contour of $-0.7$ the Hercules and Local voids become united in the region of the Serpens Caput minimum (see Table~\ref{table:minima}). 

\subsection{Sculptor Void}

\begin{figure*}[]
%\plotone{sculptor_void.ps}
%\plotone{SV_2panels.eps}
\plotone{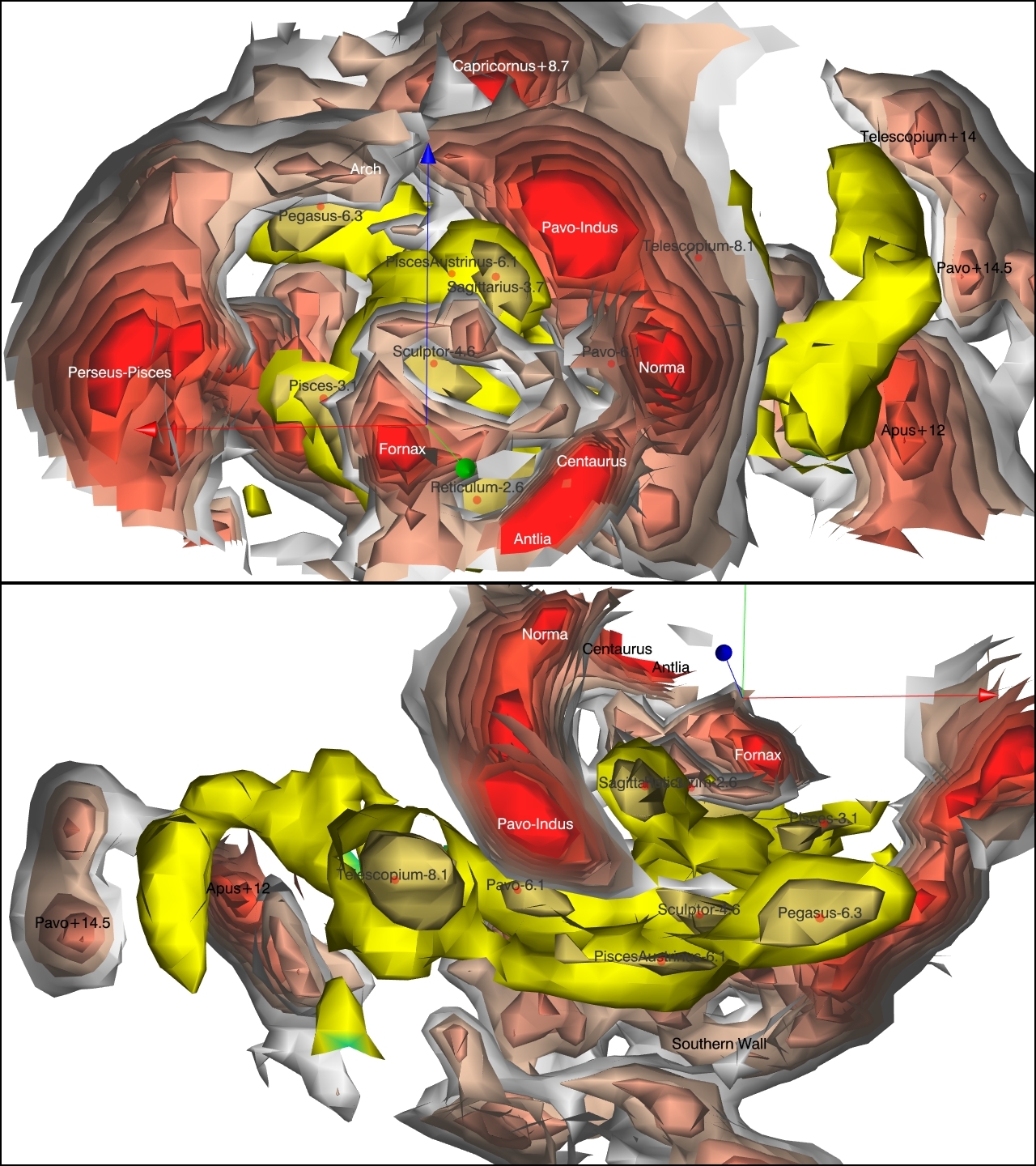}
\caption{The Sculptor Void.  Shades of yellow are used at density levels consistent with the previous 2 figures.
Once again, local minima are identified as well as prominent features on the bounding walls.
The bottom view is from near the north supergalactic pole, positive SGZ, with a foreground clip at SGZ=+3000~\kms\ to remove obstructions (08:31).  
The Southern Wall is a defining boundary at negative SGY.
In the top panel, the view is in from near the positive SGY axis.  A foreground clip at SGY=+2000~\kms\ and an extraction of the immediate area around the Virgo Cluster provides windows onto the void (08:21 and 08:49). 
}
\label{scl}
\end{figure*}

South of the galactic plane (SGY$<0$) the most adjacent dominant depression has been called the Sculptor Void \citep{1998lssu.conf.....F}.  
Its domain is illustrated in Figure~\ref{scl} and the density minima within this extensive void are listed in Table~\ref{table:minima}.  The nearest basin at density $-1.68$ lies at [$-1200$, $-1700$, $-1700$]~\kms\ $\approx$ [$-16$, $-23$, $-23$]~Mpc (Reticulum$-2.6$).  The foreground of this feature has been called the Southern Supercluster by \citet{1956VA......2.1584D} and contains the Fornax Cluster.  At the back side is the Southern Wall \citep{1990AJ.....99..751P}.    The \citet{2013AstBu..68....1E} hypervoid HV2 is coincident with the part of our Sculptor Void within 3,000~\kms.

%The deepest minimum that we record in the Sculptor Void with the CF3 reconstruction at density $-2.05$ is at the Sagittarius$-4$ location.  
Our Sculptor Void is very large, with rough dimensions at the isodensity level $-0.7$ of $\Delta$SGX,SGY,SGZ = 14500,6500,10200~\kms\ = 193,87,136~Mpc, enclosing $\sim 2 \times 10^6$~Mpc$^3$.  At isodensity level $-1.2$ the Sculptor Void already makes a link with the Local Void in the vicinity of the Pisces$-3$ minimum.  If we might be impressed that we live adjacent a void with characteristic diameter 60~Mpc, it gives perspective to know that immediately beyond there are two voids that are twice as big and with ten times greater volume.

An important entity is showing up at the outer reaches of our model that is the inner extension of what has been called the Eridanus supervoid \citep{2016MNRAS.455.1246F, 2016MNRAS.462.1882K}.  Without prejudice as to the full extent of this feature, we will call what we see the Eridanus Void.  The deepest minimum within our reconstruction is at Puppis$-6.2$ with density $-1.52$.   At its near side this void links with the Sculptor Void near the Canis Major$-4.6$ minimum.  At a distant part of the Eridanus Void, at the South Pole$-7.8$ minimum, there is no clear separation between Eridanus and Sculptor.

\begin{figure*}[]
%\plotone{3voids.ps}
\plotone{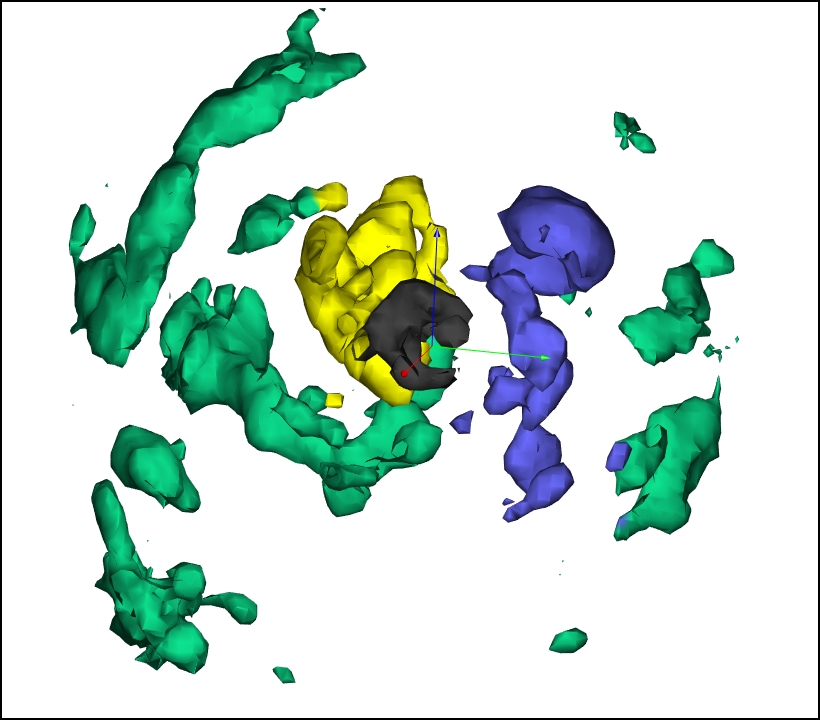}
\caption{All and only the voids.  Surfaces of all voids in the {\it Cosmicflows-3} model at the density level $-0.7$.  The Local Void is colored black, the Hercules Void is blue, the Sculptor Void is yellow and all other voids are colored green.  The view is from the reference orientation, with the Milky Way at the origin of the red, green, blue arrows (10:06).  
}
\label{3voids}
\end{figure*}

As a summary, Figure~\ref{3voids} shows the relationships between the Local, Hercules, and Sculptor voids and a glimpse of the voids further away.  The choice of a display at density $-0.7$ is arbitrary.  All these voids can be linked at values of density below the cosmic mean.  Each of the displayed voids fragment into separable pieces at more negative density cuts.  
The second interactive model\footnote{https://sketchfab.com/models/78885b3d303d4b6e99cfe099b43929fb} should be launched in order to immersively experience the panapoly of nearby voids.

\begin{table*}
\centering
\caption{Locations of density minima within the nearest voids}
\label{table:minima}
\begin{tabular}{llrrrrrrl}
\hline
   Void   &   Density &    SGX &  SGY &  SGZ &  SGX & SGY & SGZ & Description \\
\hline
       &    &   \kms    &   \kms     & \kms  & Mpc & Mpc  & Mpc &  \\
\hline
Local      & $-1.89$ &  +1650   & $-700$  &  +1650   &   +22  & $-9$ &  +22  & Lacerta$-2.4$\\
Local      & $-1.53$ &  +2100   & $-700$  &  $-300$  &   +28  & $-9$ & $-4$  & Andromeda$-2.3$\\
Local      & $-1.13$ &  $-200$  & $-200$  &   +700    &   $-3$ & $-3$ &  +9    & Aquila$-0.8$\\
Local      & $-0.93$ &  +3100   &  +1700  &  +1200   &   +41  &   +23 & +16   & UMi$-3.7$\\
\hline
Hercules & $-1.87$ & $-1200$ & +4000   &  +5000   & $-16$ & +53  &  +67   & Hercules$-6.5$\\
Hercules & $-1.67$ & $ -200$  & +6400   &  +5000   & $  -3$ & +85  &  +67  & Bo\"otes$-8$ \\
Hercules & $-1.55$ &   +2100  &  +3500   & $-1200$ &   +28 & +47   & $-16$ & UMa$-4.3$; Lower Arrowhead\\
Hercules & $-1.39$ & $-3100$ &  +4000   & $-5400$ & $-41$ & +53   & $-72$ & Sextans$-7.4$\\ 
Hercules & $-1.30$ &  $-200$  &  +5000   & $-1700$ &  $-3$ & +67   & $-23$ & Leo$-5.2$; Coma\\
Hercules & $-1.28$ & $-1200$ & +3100   &  +2100   & $-16$ & +41  &  +28   & Serpens Caput$-3.9$\\
Hercules & $-1.12$ &   +1200   &  +5400   & $-5000$ &    +16  & +72    & $-67$ & Leo$-7.5$\\
Hercules & $-1.00$ &   +3500  &  +4500   &   +700   &   +47  & +60   &    +9   & UMa$-5.8$; Far Arrowhead\\
%Hercules & $-1.1  $ &   +2400  &  +3700   &  +3300   &   +32  & +49   &  +44   & Upper Arrowhead Void \\
\hline
Sculptor  & $-1.68$ & $-1200$ & $-1700$ & $-1700$ & $-16$ & $-23$ & $-23$ & Reticulum$-2.6$\\
Sculptor  & $-1.55$ & $-1700$ & $-2100$ & $+2600$& $-23$ & $-28$ & $+35$ & Capricornus$-3.7$\\
Sculptor  & $-1.53$ & $+1700$ & $-2600$ & $+200$ & $+23$ & $-35$ & $+3$ & Pisces$-3.1$\\
Sculptor  & $-1.43$ & $+1700$ & $-4500$ & $+4000$ & $+23$ & $-60$ & $+53$ & Pegasus$-6.2$\\
Sculptor  & $-1.39$ & $-6400$ & $-4000$ & $+3000$ & $-85$ & $-53$ & $+40$ & Telescopium$-8.1$\\
Sculptor  & $-1.32$ & $+2600$ & $-2100$ & $+300$  & $+35$ & $-28$ & $+4$   & Pisces$-3.4$\\
Sculptor  & $-1.23$ & $-1200$ & $-5400$ & $+2600$ & $-16$ & $-72$ & $+35$ & Pisces Austrinus$-6.1$\\
Sculptor  & $-1.20$ & $-4500$ & $-4100$  & $+700$   & $-60$ & $-55$ & $+9$ & Pavo$-6.1$\\
Sculptor  & $-1.19$ & $-700$    & $-4500$ & $+700$   & $-9$  & $-60$ & $+9$   & Sculptor$-4.6$\\
\hline
Eridanus & $-1.52$ & $-3100$ & $-2100$ & $-5000$ & $-41$ & $-28$ & $-67$ & Puppis$-6.2$\\
Eridanus & $-1.49$ & $ -200$ & $ -700$ & $-4500$ & $-3$   & $-9$   & $-60$ & Canis Major$-4.6$\\
Eridanus & $-1.38$ & $+1100$ & $-7400$ & $-2200$ & $+15$ & $-99$ & $-29$ & Cetus$-7.8$\\
Eridanus & $-1.08$ & $-6000$ & $-2600$ & $-3600$ & $-80$ & $-35$ & $-48$ & Chamaeleon$-7.4$\\
\hline
Other      & $-1.17$ & $+700$  & $-8700$ & $+9200$ & $+9$  & $-116$ & $+123$ & Aquarius$-12.7$\\
Other      & $-1.07$ & $-8200$  & $-10100$ & $+2800$ & $-109$  & $-135$ & $+37$ & Indus$-13.3$\\
Other      & $-0.78$ & $-12500$  & $-4600$ & $-6800$ & $-167$  & $-61$ & $-91$ & South Pole$-15$\\
\hline
\hline
\end{tabular}
\end{table*}

\section{Morphology of the Walls}

A detailed discussion of the over dense regions will be left to another day, but we give attention here to the immediate walls around the Local, Hercules, and Sculptor voids.

With the Local Void, the dominant bookend bounding features are the Perseus-Pisces complex at SGX$\sim+4500$~\kms\ \citep{1988lsmu.book...31H} and at SGX$\sim -4000$~\kms\ the Pavo-Indus arm rising out of the region called the Great Attractor through the Norma Cluster \citep{1987ApJ...313L..37D, 1996Natur.379..519K}.  At right angles, the most prominent features at positive and negative supergalactic latitudes are the Arch at SGZ$\sim+4200$~\kms\ and the Centaurus-Puppis-PP filament at SGZ$\sim -2700$~\kms\ \citep{2017ApJ...845...55P}.  These ceiling and floor of the Local Void are poorly documented in redshift surveys because of galactic obscuration but coherent velocity patterns provide robust reconstructions.

The orthogonal directions of $\pm$SGY, by contrast, are toward the uncontaminated and well observed galactic poles.  In the galactic north the Local Void is bounded by de Vaucouleurs' Local Supercluster at SGY$\sim+2500$~\kms\ running through the Virgo and Centaurus clusters \citep{1956VA......2.1584D}.  For all the importance we have given this structure it is not very substantial.  The Local Void easily makes connections through this region with the Hercules Void.  The Local Void limits are even more porous at negative SGY.  De Vaucouleurs' Southern Supercluster in Fornax and Eridanus and the strand extensions described by \citet{2013AJ....146...69C} are the weak separators from the Sculptor Void at SGY$\sim-2500$~\kms.
We will return later to a discussion of wispy galaxy filaments bounding and permeating the Local Void.

Turning attention to the Hercules Void, dominant structures on the back side are the Great Wall at SGY$\sim +7000$~\kms\ \citep{1986ApJ...302L...1D} merging into the Hercules complex at SGZ$\sim +7000$~\kms\ \citep{1984ApJ...277...27B} and further merging into the Ophiuchus \citep{1981ApJ...245..799J} and Libra+8 structures at SGX$\sim-6000$~kms.  On the near side, the over densities are the less consequential Local Supercluster, creating a separation from the Local Void, and such other minor features as the Hydra$-$Cancer filament and the Arrowhead mini-supercluster \citep{2015ApJ...812...17P}.  Above the supergalactic equator (SGZ$>0$) essentially everywhere locally densities are below the mean.  It is through this space that the Hercules and Local voids connect.  \citet{1995ApJ...454...15S} have recorded the kinematic manifestation of this general under density in the ubiquitous flow toward negative SGZ of nearby galaxies (see Figure~\ref{orbits} and related interactive model and video sequence).
At present, the full Hercules Void is poorly constrained on the +SGX side.  {\it Cosmicflows-4}, the next edition of our catalog of distances will provide more satisfactory coverage of this part of space. 

On the other side of the sky, the Sculptor Void is held in the far side embrace of the Southern Wall \citep{1990AJ.....99..751P} running at roughly SGY$\sim-7000$~\kms\ from the Perseus-Pisces region all the way to structure at the celestial South Pole.  This latter feature appears to be very important and we expect to discuss it in detail in a future paper.  The nearer side of the Sculptor Void is bounded by the minor structures on the negative SGY side of the Local Void discussed above.   There is easy penetration between the Sculptor and Local voids where de Vaucouleurs' Southern Supercluster peters out beyond the Fornax-Eridanus complex.  In detail, we see the Southern Wall as bifurcating into what we call the North Fork (Pegasus+8.5 filament) and the South Fork (Grus-Pisces Austrinus+10 filament).  The North Fork connects through Capricornus+7 to Ophiuchus forming a roof over the Local Void extending to above the Hercules Void.  At large values of negative SGX and SGY the Sculptor Void boundaries are at the challenging limits of our reconstruction and dissolve in places into what we call the Eridanus Void.

\section{The V-Web Representation of Voids}

The construction of structure up to this point in the discussion have been based on a model of the density field derived from the divergence of the three-dimensional velocity field in accordance with linear theory.  An alternative representation is derived from a calculation of the shear of the velocity field at a given location \citep{2012MNRAS.425.2049H}.
\begin{equation}
\Sigma_{\alpha\beta} = -  ( \partial_\alpha v_\beta  +  \partial_\beta v_\alpha)/2H_0
\label{eq;vweb}
\end{equation}
where partial derivatives of the velocity ${\bf v}$ are determined along directions $\alpha$ and $\beta$ of the orthogonal supergalactic Cartesian axes, normalized by the average expansion given by the Hubble Constant, $H_0$.  Eigenvalues indicating collapse have negative values. 

The eigenvectors of the shear define the principal axes of collapse and expansion.  Knots, filaments, sheets, and voids are associated, respectively, with 3, 2, 1, and 0 positive eigenvalues.  These four domains can be separated by surfaces of the eigenvalues.  We refer to such representations as the cosmic velocity (V) web \citep{2017NatAs...1E..36H, 2017ApJ...845...55P}.

\begin{figure*}[!]
\centering
%\special{psfile=vweb_2panels.ps hscale=32 vscale=32 voffset=-550 hoffset=120}
%\plotone{vweb_2panels.ps}
%\plotone{SDvision_V-Web_Voids3components_view12_v004.jpg}
\includegraphics[scale=0.4]{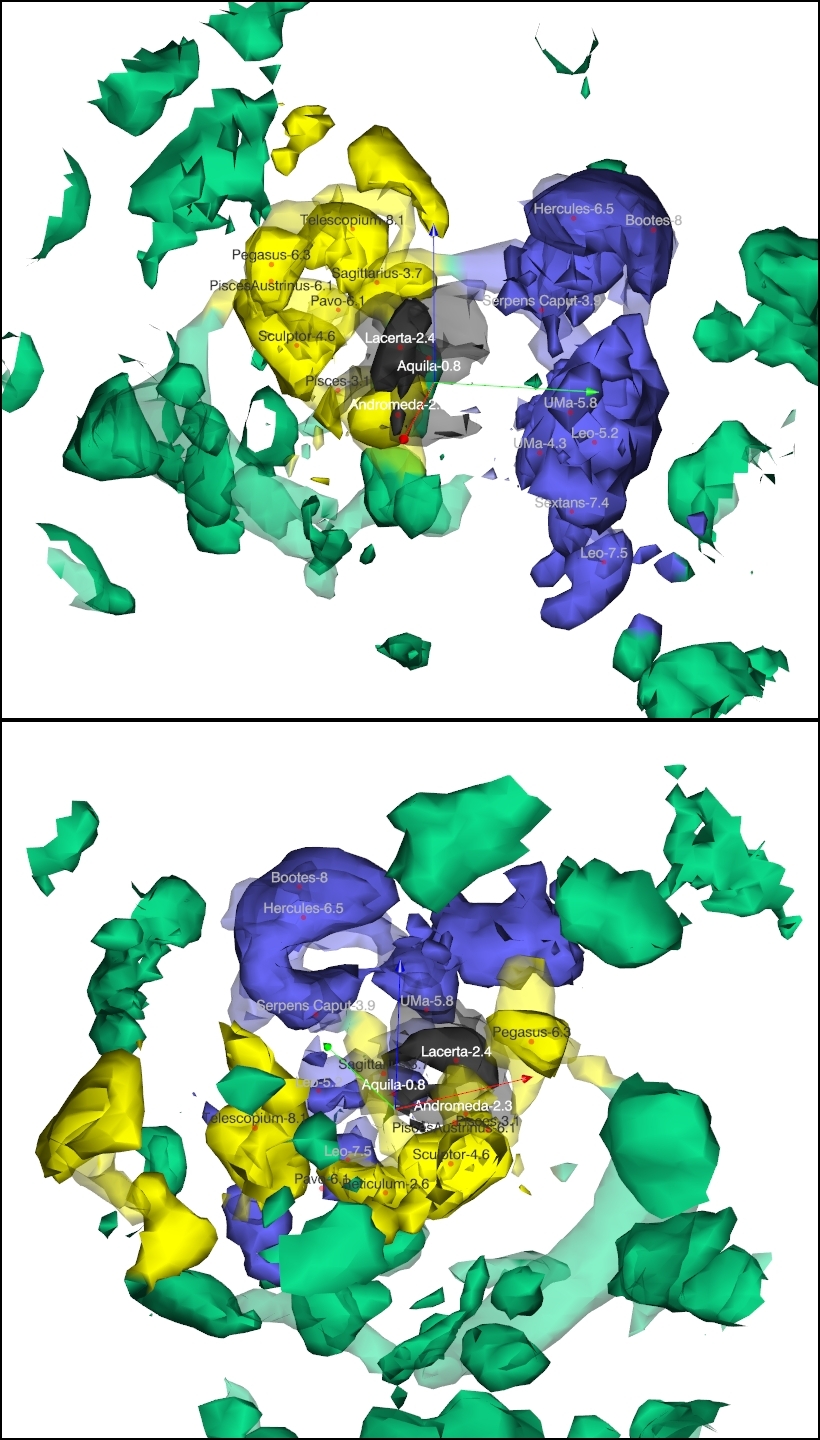}
\caption{V-web representation of voids.  Voids (expansion on 3 cardinal axes) are represented by solid surfaces, with the Local Void in black, the Hercules Void in blue, the Sculptor Void in Yellow, and other voids in green.  Sheets (expansion on 2 axes, collapse on the third) are represented by transparent surfaces at an arbitrary eigenvalue.  Locations and names of deepest density troughs are carried over from previous figures.  In the top panel, the view is from the reference direction while in the lower panel the scene has been rotated to a view from negative SGX, SGY, positive SGZ.
}
\vspace{18cm}
\label{vweb}
\end{figure*}

The current interest is in the voids, locations with expansion along all three axes.  Figure~\ref{vweb} presents an alternate to the density isocontour plot of Figure~\ref{3voids}; the V-web representation of voids and sheets.  Here, regions with expansion on three axes (voids) are shown as solid colors, consistent with the schema in previous figures, while regions with expansion on only two axes (sheets) are shown by transparent surfaces in related colors.  
We have chosen an arbitrary eigenvalue level for the display of the sheet isosurface that roughly parallels the arbitrarily chosen density isocontours of previous figures.

The alternative V-web and isodensity representations are similar (of course they are drawn from the same data and analysis) but there are curious differences.  The deepest basins have the same locations.  However it is interesting as an example to give attention to the sheet-topology link between the Hercules and Sculptor voids bypassing the Local Void seen in the top panel of Fig.~\ref{vweb} (near the feature named Sagittarius$-3.7$).   It is found that minor filaments separate the Local Void from a tunnel connecting the Sculptor and Hercules voids.

\section{Structure Defined by a Redshift Survey}

It is worth briefly to compare structural features defined by {\it Cosmicflows-3} velocities with the redshift space distribution of galaxies. 
The current interest is in underdense regions where there are relatively few galaxies.
We give attention alternatively to the walls that bound voids and to the minor strands of galaxies that can permeate voids and give separation to adjacent minima.

Our comparisons are made with the redshift compilation V8k that was described by \citet{2012ApJ...744...43C}.  The sample consists of 30,124 galaxies within a cube that extends from the origin $\pm8000$~\kms\ on the cardinal axes in supergalactic coordinates.  This sample covers the entire sky reasonably uniformly except at the Galactic plane and provides relatively dense coverage locally where we can most meaningfully make comparisons.

A wide angle comparison between the V8k redshift sample and the iso-density outline of the Local Void can be seen in the accompanying video in frames following 03:28.  Almost all the individual galaxies lie outside the contours of the void although several filamentary features adhere closely to the void boundaries (emphasized transiently in blue in the video).  The few exceptions where filaments penetrate the void are worth commentary.

\begin{figure*}[!]
%\begin{center}
\centering
%\special{psfile=PerseusCloud-MW_posX.ps hscale=30 vscale=30 voffset=-320 hoffset=30}
%\special{psfile=PerseusCloud-MW_negY.ps hscale=30 vscale=30 voffset=-530 hoffset=30}
%\special{psfile=NGC1161.ps hscale=45 vscale=45 voffset=-370 hoffset=375}
%\special{psfile=NGC1169.ps hscale=45 vscale=45 voffset=-270 hoffset=350}
%\vspace{200mm}
%\plottwo{PerseusCloud-MW_posX.png}{PerseusCloud-MW_negY.png}
%\includegraphics[width=\textwidth]{LocalVoid_PerseusCloud_view12_v001.jpg}
\includegraphics[width=\textwidth]{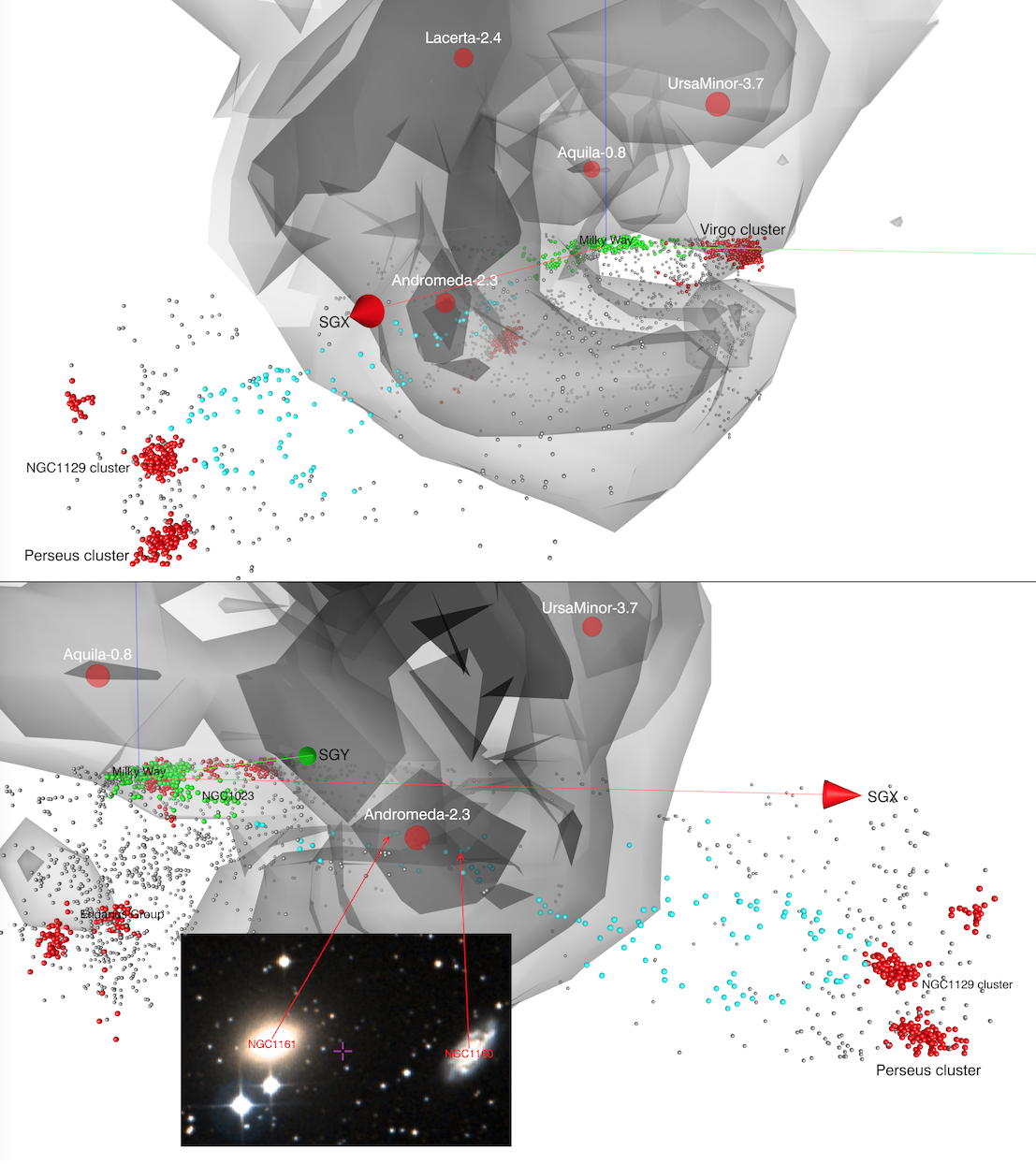}
\caption{Two rotated views of the Perseus Cloud filament passing from the Virgo Cluster, past the Milky Way, through the deep Local Void minimum of Andromeda-2.3, to the vicinity of the Perseus Cluster.  The image in the inset is of the giant lenticular galaxy NGC~1161, and its spiral companion NGC~1160, deep within the Local Void. (video frames 04:44 to 05:21)}
\vspace{18cm}
%\end{center}
\label{perseus}
\end{figure*}

A very sparse filament dramatically spans between the Virgo Cluster and the Perseus Cluster in a very direct route that takes it through a deep minimum in the Local Void.  The commencement is the Local Sheet that includes the Milky Way and extends from  the proximity of Virgo to the NGC~1023 Group.  In the Nearby Galaxies Atlas \citep{1987nga..book.....T} the continuation is called the Perseus Cloud, basically receiving a new name just because of the tenuousness and severe obscurity of the structure. The Nearby Galaxies Atlas sample cuts off at 3,000~\kms\ but the V8k sample reveals the extension all the way to the vicinity of the Perseus Cluster.  This wispy filament is isolated in the two panels of Figure~\ref{perseus}.  Remarkably, it passes very close to the deep density minimum Andromeda$-2.3$ and it is very near this location that we find a very substantial grouping of galaxies dominated by the lenticular NGC~1161 (see inset within Fig.~\ref{perseus}).  This S0 galaxy has near infrared luminosity log$L_K = 11.36$ assuming a distance of 32~Mpc.  At least three more large galaxies with log$L_K > 11$ reside in the vicinity: NGC~1169, NGC~1186, and PGC~11586.

\begin{figure*}[!]
%\special{psfile=PegasusCloudSpur_negXY.ps hscale=22 vscale=22 voffset=-250 hoffset=70}
%\special{psfile=PegasusCloudSpur_posYZ.ps hscale=25 vscale=25 voffset=-590 hoffset=70}
%\vspace{220mm}
%\plottwo{PegasusCloudSpur_negXY.png}{PegasusCloudSpur_posYZ.png} 
\includegraphics[width=\textwidth]{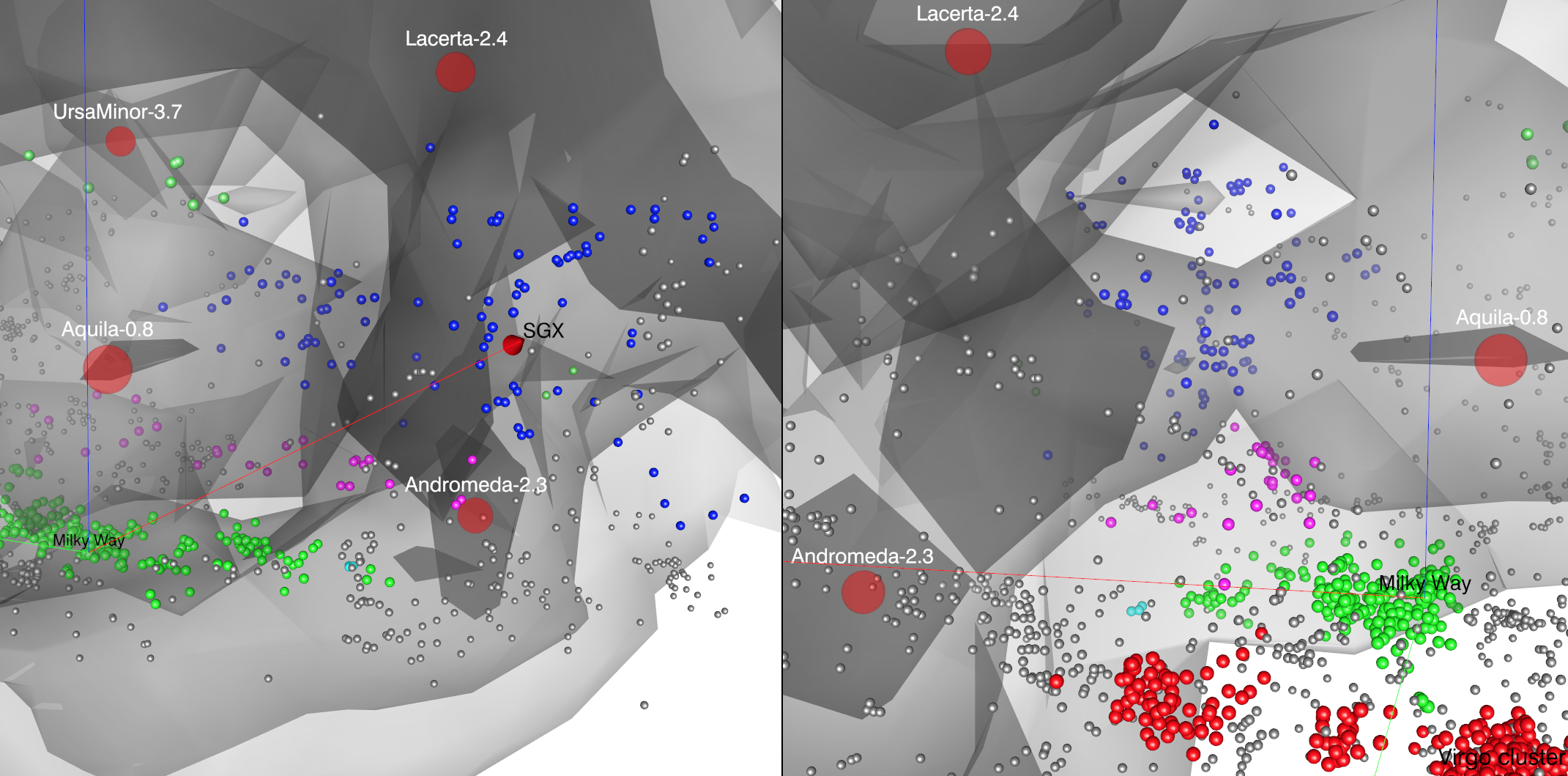}
\caption{Two views of the filaments Pegasus Cloud (galaxies in blue) and Pegasus Spur (galaxies in magenta) that thread through the Local Void.  The Pegasus Cloud penetrates the Local Void between the Lacerta$-2.4$ and Aquila$-0.8$ minima while the Pegasus Spur wraps closely on the underside of Aquila$-0.8$. (video frames 05:23 to 05:42)}
\label{pegasus}
\end{figure*}

The second case that draws our attention involves entities called the Pegasus Cloud and Pegasus Spur in the Nearby Galaxies Atlas.  Associated galaxies are illustrated in Figure~\ref{pegasus}.  The Pegasus Spur lies closely outside the mid iso-density contour confining the Local Void.  The Pegasus Cloud is particularly interesting because it coincides with a higher density `tunnel' between the abysses Lacerta$-2.4$ and Aquila$-0.8$. 

%\clearpage

%\clearpage

\section{Consequences of the Local Void on our Motion}

It is to be noted that Lacerta$-2.4$, the deepest part of the Local Void, is aligned with the anti$-$apex of the cosmic microwave background dipole.   \citet{2017NatAs...1E..36H} have already brought attention to the likely importance of a very large underdensity (the Dipole Repeller) in generating the 631~\kms\ motion of the Local Group with respect to the microwave background frame.  We suggest that the Local Void makes a significant additive contribution.  

The apex of the Local Group motion in supergalactic coordinates is toward $SGL=139$, $SGB=-31$.  The major influences that give rise to this motion can be separated into nearby, intermediate, and far domains.  We will use two distinct approaches to evaluate the sources of our deviant motion.

We begin with the first of these.
The nearby domain was studied in detail by \citet{2017ApJ...850..207S} with numerical action orbit reconstructions, illustrated in Figure~\ref{orbits} and following 05:50 in the accompanying video.  This domain extends to 38 Mpc $\sim2850$~\kms\ and includes the traditional Local Supercluster dominated by the Virgo Cluster.  It excludes the so-called Great Attractor region, the densest part of our Laniakea Supercluster.  Structure beyond 38~Mpc was represented in the numerical action model by external tidal fields as given by a Wiener filter linear theory rendition based on {\it Cosmicflows-2} distances \citep{2014Natur.513...71T}.

Taking the average of the orbits of the Milky Way and M31, we find a Local Group SGX, SGY, SGZ motion with respect to the center of mass within 38~Mpc to be [$-105$, 314, $-192$]~\kms.  It can be seen from the position of the Local Group within the embrace of the Local Void (Figures \ref{LV} and \ref{orbits}) that the velocities toward negative SGX and SGZ can be attributed in large measure to repulsion from the Local Void.  The amplitudes are consistent with values determined locally \citep{2019ApJ...872L...4A} (Anand et al. 2019b in press).

The Local Void might contribute to the positive SGY motion but in this direction the Virgo overdensity must dominate.  In the Shaya et al. model the spherical volume centered on the Virgo Cluster extending to the Local Group is a factor 1.39 above mean density.  In the spherical approximation, this overdensity would attract us at $\sim300$~\kms.  Tidal squeezing (from a filament running into Virgo parallel to SGX) and distending (from voids at $\pm$SGZ) are complications to this estimate.  In the numerical action model the differential Virgo $-$ Local Group velocity is 200~\kms.  In this model, the SGY motion in the near region of $\sim +300$~\kms\ is roughly the sum of  a pull of $\sim200$~\kms\ from the Virgo overdensity and  a push of $\sim 100$~\kms\ from the Local Void.

The intermediate domain, from 38~Mpc to roughly 100~Mpc is dominated by the competition between the Laniakea and Perseus$-$Pisces attractors. The core of Laniakea (the Great Attractor) contains the clusters Centaurus, Norma, Hydra, Pavo~II, A3365, A3537, A3574, and S753.  Tugging in the opposite direction are the chain of clusters including Perseus, Pisces, A262, A347, and NGC~507.  Laniakea is clearly winning in influence at our position.  To a reasonable approximation, it is the intermediate domain that dominates the tidal forces on the inner 38~Mpc zone as calculated from the Wiener filter model based on {\it Cosmicflows-2} distances.  The bulk motion of the inner region due to these mostly intermediate zone influences has the SGX, SGY, SGZ vector components [$-212$, 95, $-106$]~\kms\ \citep{2017ApJ...850..207S}.   The SGX and SGY components reflect the competition between Laniakea and Perseus$-$Pisces while the SGZ component reflects the great extent of the Local Void, reaching well beyond the 38~Mpc limit of the near region.

The far domain must account for the remainder.  The sum of the inner and intermediate zones produce Local Motion of [$-334$, 411, $-296$]~\kms, leaving a residual of [$-76$, $-58$, $-28$]~\kms, attributable to a distant pull from the Shapley Concentration and push from the Dipole Repeller \citep{2017NatAs...1E..36H}.  The separation between the intermediate and far domains is approximate.  Part of the attribution to the intermediate domain may arise from the far domain.

\begin{table*}
\centering
\caption{Sources of Local Group Motion}
\label{table:shells}
\begin{tabular}{lcccc}
\hline
Numerical Action Analysis & & & & \\
\hline
   Zone & SGX & SGY & SGZ & Sum \\
\hline
        & \kms& \kms& \kms & \kms \\
\hline
Near ($<38$~Mpc)     & $-122$ &  316  & $-190$ & 388 \\
Mid (38 $-$ 100~Mpc) & $-212$ &   95  & $-106$ & 255 \\
Far ($>100$~Mpc)     & $ -76$ & $-58$ & $ -28$ & 100 \\
\hline
Cumulative           & $-410$ &  353  & $-324$ & 631 \\
\hline
Wiener Filter Analysis & & & & \\
\hline
Local Void           & $-129\pm59$ & $43\pm35$  & $-142\pm36$ & $197\pm77$ \\
Greater Virgo        & $22\pm50$   & $281\pm38$ & $-8\pm22$   & $282\pm66$ \\
LV+Virgo             & $-148\pm58$ & $329\pm42$ & $-150\pm37$ & $391\pm80$ \\
Full WF              & $-399\pm49$ & $355\pm32$ & $-337\pm42$ & $632\pm72$ \\
\hline
\hline
\end{tabular}
\end{table*}

Uncertainties in the one-dimensional components of velocity deviations in these various ranges are estimated to be at the level of $\pm40$~\kms.  The estimate is approximate because our break-out of influences is approximate.  Our motion is the product of innumerable actors.  As for the influence of the Local Void, is it that we are participating in the void expansion or responding to multiple ovedensities outside the void?  Whatever the semantics, the numerical action model predicts that our proximity to the Local Void results in a deviant motion of roughly [$-100$, 100, $-200$]$\sim250$~\kms.

Alternatively, the impact of respectively the Local Void and the greater Virgo Cluster can be evaluated from the Wiener filter model.  In the case of the Local Void, the influence at our position is found summing over the volume defined as the Local Void below the contour of $\delta=0$.  In the case of the Virgo Cluster, the summation is over a sphere centered on the cluster extending in radius to our position.  Statistical uncertainties are determined by averaging over multiple constrained realizations \citep{1991ApJ...380L...5H, 1999ApJ...520..413Z}.  The numerical results are gathered in Table~\ref{table:shells} along with those from the numerical action analysis.

The most directly comparable results between the two analyses (besides the cumulative values) are the numerical action ``near'' row and the Wiener filter ``LV+Virgo'' row.  The results are in statistical agreement.  The comparison is approximate since in detail the contributions are not the same.  Give consideration to the distinct SGX, SGY, SGZ components.  There is agreement that, in sum, the Virgo overdensity and Local Void are combining to give us a deviant motion of about 300~\kms\ toward positive SGY with Virgo dominant at the level of $65-80\%$.  The Local Void is held responsible for the substantial deviant motions toward negative SGX and SGZ.

Overall, roughly, 50\% of our motion reflected in the cosmic microwave background dipole fluctuation is taken to arise relatively nearby.  The Local Void is a major contributor.  From the breakdown presented in Table~\ref{table:shells}, it can be inferred that the Local Void is a dominating influence on motions in the (negative) SGZ direction (appreciating that the Local Void extends well into the 'mid' zone), contributes at the level of 30\% in the (negative) SGX direction, and contributes modestly in the (positive) SGY direction.  These conclusions are consistent with the prescient early claims by \citet{1988MNRAS.234..677L} and \citet{1988lsmu.book..199L}.

\section{Final Thoughts}

The model of nearby structure in the universe that has been presented is derived strictly from the radial velocities of test particles assuming deviations from uniform expansion arise from Newtonian gravity in a basic $\Lambda$CDM cosmology.  There is a manifestly reasonable agreement with the alternative of structure inferred from redshift surveys.  Much remains to give the comparison a quantitative foundation.  We have introduced the morphology of structure determined from the 18,000 distances of {\it Cosmicflows-3} with a study of voids because such regions are simpler than high density regions.  The reconstruction by \citet{2019arXiv190101818G} has relatively coarse resolution of $6.25/h_{75}$~Mpc.  Higher resolution can be achieved at the expense of computation time.   Studies of high density regions will benefit from high spatial resolution where the quasi-linear regime can be probed with the techniques discussed by \citet{2018NatAs...2..680H} and numerical action methods can probe the fully non-linear regime \citep{2017ApJ...850..207S}.  

Even if voids are simpler, with dynamics in the linear regime, their morphologies have been much less well understood than that of high density structures.  Yet we live at the edge of a void.  Nearby our measurements of distances of galaxies are plentiful and associated peculiar velocities are well determined.  This information, processed through the Wiener filter, allows us to define the properties of the Local Void with considerable precision even where much information is lost due to Galactic obscuration.

The Local Void does not have a simple shape.  Now, for the first time, we have a map that reveals its complexity.   It is unambiguously demonstrated that the Local Void, familiarly prominent at positive SGZ, and the void in front of the Perseus-Pisces filament \citep{1986ApJ...306L..55H} at positive SGX are parts of the same extensive under density.  This linkage has not been evident because of intervening obscuration.

Attention has been drawn to the substantial contribution of the Local Void to our motion reflected in the cosmic microwave background dipole anisotropy.  The Local Group has a deviant motion due to the Local Void of $200-250$~\kms\ with respect to the center of mass within 38~Mpc (and another $200-250$~\kms\ due to the Virgo overdensity).  The full effect of the Local Void as it extends to the zone beyond the 38~Mpc near region has a repelling influence that explains most of the SGZ component of our motion in the cosmic microwave background frame.  What is left over after accounting for these nearby actors is $\sim300$~\kms, directed mostly toward negative SGX, and attributable to structure in the mid and far shells discussed in the previous section.  

%It is to be noted that Lacerta$-2.4$, the deepest part of the Local Void, is aligned with the anti$-$apex of the cosmic microwave background dipole.   \citet{2017NatAs...1E..36H} have already brought attention to the likely importance of a very large underdensity (the Dipole Repeller) in generating the 630~\kms\ motion of the Local Group with respect to the microwave background frame.  We suggest that the Local Void makes a significant additive contribution.  In roughly comparable fractions, our peculiar motion is predominantly due to the fortuitous alignment of attractions from the Centaurus-Hydra-Norma (Great Attractor) and Shapley over densities in one direction and the Dipole Repeller and Local Void under densities in the opposite direction. 

We have revisited the known fact \citep{1995A&A...301..329L} that chains of galaxies can thread through voids.  Usually the constituents are small galaxies.   The case we present of the Perseus Cloud that traverses the Local Void from the Virgo Cluster to the Perseus Cluster is particularly noteworthy, not only because we are part of it.  Remarkably, it passes through one of the lowest density parts of the Local Void and near that location is a significant gathering of substantial galaxies.  Regrettably, these systems lie at a low Galactic latitude and have been poorly studied.  

Immediately beyond the Local Void lie two much bigger under dense regions that we call the Hercules Void and the Sculptor Void.   In fact the voids are all interconnected by necks that are below the mean density.  Boundaries can be arbitrary and will undoubtedly be sources of contention.  We use our discoverer's prerogative and give names to outstanding features.  

Our cartography reveals hints of the complexity of overdenities in the region we are exploring within $0.05c$.  Even our discussion of voids represents only a first pass.  There is much more to be learned as the density of data becomes richer and we gain confidence in reconstructions with increasing resolution.  

\bigskip
Support for this program is provided by Grant No. 80NSSC18K0424 from the US National Aeronautics and Space Administration and from multiple awards from the Space Telescope Science Institute, most recently HST-AR-14319, HST-GO-14636, and HST-GO-15150.  HC acknowledges support by the Institut Universitaire de France and the CNES.

\clearpage
%\begin{figure*}[]
%\plottwo{PegasusCloudSpur_negXY.ps}{PegasusCloudSpur_posYZ.ps}
%\caption{Structure within voids.  This view is from near the negative SGY axis, with a cut of structure at SGY$<-2500$~\kms.
%}
%\label{PegasusCloud}
%\end{figure*}

%\begin{figure*}[]
%\plotone{redshifts_LV_SGXle1000.ps}
%\caption{Structure within voids.  In this view from positive SGX, structure at SGX$>+1000$~\kms\ is removed to provide a window across (from left to right) the Sculptor, Local, and Hercules voids.  Galaxies from the V8K sample are located by white dots.  A density surface at ??? provides walls around minor filaments.
%}
%\label{redshifts1}
%\end{figure*}

%\begin{figure*}[]
%\plotone{redshifts_LV_SGYge-2500.ps}
%\caption{Structure within voids.  This view is from near the negative SGY axis, with a cut of structure at SGY$<-2500$~\kms.
%}
%\label{redshifts2}
%\end{figure*}

\bibliography{paper}
\bibliographystyle{aasjournal}

%\onecolumn
%\begin{figure}[htbp]

%\twocolumn

\end{document}